\documentclass[%
reprint,
superscriptaddress,
nofootinbib,
amsmath,
amssymb,
aps,
floatfix,
showkeys,
]{revtex4-2}

\usepackage[colorlinks,citecolor=blue,linkcolor=blue,anchorcolor=blue,filecolor=blue,
urlcolor=blue]{hyperref}

\usepackage{graphicx,color,xcolor}
\usepackage{amsfonts,amsmath}
\usepackage{amssymb,ulem}
\usepackage{dcolumn}
\usepackage{bm,lipsum}
\usepackage[utf8]{inputenc}
\usepackage{subeqnarray}
\usepackage{array}
\usepackage{epsfig,hyperref}
\usepackage{morefloats}
\usepackage{relsize}
\usepackage{url}

\usepackage{comment}

\usepackage{cleveref}
\usepackage{orcidlink}

\begin{document}

\preprint{ }

\title{XTE J1814-338 as a dark matter admixed neutron star}

\author{Luiz L. Lopes}
\email{llopes@cefetmg.br}

\affiliation{ Centro Federal de Educac\~{a}o Tecnol\'{o}gica de Minas Gerais Campus VIII; \\ CEP 37.022-560, Varginha - MG - Brazil\\}

\author{Adamu Issifu}
\email{ai@academico.ufpb.br}

\affiliation{Departamento de F\'isica e Laborat\'orio de Computa\c c\~ao Cient\'ifica Avan\c cada e Modelamento (Lab-CCAM), Instituto Tecnol\'ogico de Aeron\'autica, DCTA, 12228-900, S\~ao Jos\'e dos Campos, SP, Brazil}

\date{\today}

\begin{abstract}
The existence of the ultracompact object XTE J1814-338, with an inferred mass and radius of $M$ = 1.21 $\pm$ 0.05 $M_\odot$ and R = 7.0 $\pm$ 0.4 km, presents a great challenge for the theory of neutron stars. Within this context, we revisit the theory of dark-matter-admixed neutron stars and infer the physical properties of this compact object, such as the Fermi momentum of dark matter necessary to compress the star at such low radius, its equation of state, speed of sound, and some macroscopic properties, such as the moment of inertia and the dimensionless tidal parameter. We also compare the physical properties of the XTE J1814-338 with other pulsars, such as the canonical 1.4 M$_\odot$, the PSR J0740 + 6620, and the HESS J1731-347.  


\end{abstract}

\maketitle

\section{Introduction}
The study of particle composition in the interior of neutron stars (NSs) is an open problem that is actively being researched in dense nuclear matter physics. The most obvious possibility is that
NSs are composed of pure hadronic degrees of freedom (nucleons and exotic baryons) \cite{Ribes:2019kno, debora-universe}. However, the extreme conditions encountered in its core, theoretically permit the dissociation of baryons into free quarks in the interior (hybrid stars), perhaps, through a first-order phase transition \cite{Cierniak:2021knt, Heiselberg:1999mq, Lopes_2022, Issifu:2023ovi, Fukushima:2010bq}. In this case, a quark core surrounded by a hadronic matter is expected. This stellar construction shows good agreement with the gravitational wave constraints and recent Neutron Star Interior Composition Explorer (NICER) data \cite{Contrera:2022tqh, Cierniak:2021knt, Kumar:2023lhv}. Generally, the presence of additional degrees of freedom in the stellar core affects the physical and internal structures of the star. For instance, the presence of hyperons softens the equation of state (EoS), reducing the star's maximum gravitational mass and affecting its radius \cite{Glendenning:1991es, Weissenborn:2011kb}. Similarly, hybrid stars with a larger quark core have a softer EoS, resulting in a smaller gravitational mass and radius \cite{Lopes:2020rqn}. Consequently, the presence of dark matter (DM) in the stellar core is also expected to reduce the size of the star as discussed in \cite{Das:2021hnk, Xiang:2013xwa, Lenzi:2022ypb, Lourenco:2022fmf}

The recent work of Kini {\it et al.} \cite{Kini:2024ggu}, using pulse profile modeling technique to investigate the thermonuclear burst variations from accreting NSs has revealed a striking mass (M) and radius (R) measurement for XTE J1814-338 pulsar. The authors measured $M = 1.21^{+0.05}_{-0.05}\rm M_\odot$ and $\rm R = 7.0^{+0.4}_{-0.4}km$ at 68\% confidence level (CL). The inner composition of such an ultra-compact object has been a subject of intense research with some of them speculating that the presence of DM cannot be ruled out. Particularly, some study points to a bosonic DM star with a nuclear matter core \cite{Pitz:2024xvh} whilst others point to a strange quark star admixed mirrored DM \cite{Yang:2024ycl}. Notably, the subject of ultra-compact objects was already being investigated (see \cite{Li:2022ivt} and references therein) before the work of Kini {\it et al.}. Specifically, it was demonstrated that a hybrid EoS consistent with GW170817 and NICER measurement constraints could produce a star with extremely small radii. In \cite{Laskos-Patkos:2024fdp}, the authors established that a strong phase transition from hadronic to quark matter phase could explain the existence of such ultra-compact stars.

In this work, we investigate the role of fermionic DM inside an NS, which could lead to the possible formation of ultra-compact objects, such as XTE J1814-338. The DM model consists of dark fermions interacting with nuclear matter via the Higgs channel \cite{Das:2018frc}. As the Higgs bosons are known to interact strongly with more massive particles than lighter ones, a light neutralino with mass $M_\chi = 200\rm GeV$ is employed as a dark matter candidate to describe the dark fermions \cite{Panotopoulos:2017idn, Martin:1997ns}. Direct dark matter detection experiments such as LUX \cite{daSilva:2017swg} and XENON \cite{XENON:2017vdw} strongly exclude DM-nucleon cross section above $\sim 8\times 10^{-47}\rm cm^2$ for dark matter with masses ranging from $30-50~\rm GeV$ at 90\% CL. Moreover, the invisible Higgs decay width strongly constrains DM with masses below $m_H/2$ (with $m_H$ the mass of the Higgs boson). Therefore the choice of $m_H$ is carefully made to evade these constraints. As discussed in Refs, several authors have used similar DM models to investigate neutron star properties~\cite{Das:2018frc, Lenzi:2022ypb, Lourenco:2022fmf, Das:2020vng, Lourenco:2021dvh, Lopes:2023uxi}. 

The nuclear matter, on the other hand, is described by the quantum hadrodynamics (QHD) model~\cite{Serot_1992} with non-linear massive mesons  ($\sigma\omega\rho\phi$) mediating the interaction between the baryons~\cite{IUFSU, Miyatsu2013}. The nuclear matter is composed of the baryon octet, including both nucleons and hyperons. The DM interacts with the nuclear matter via the Higgs portal as mentioned above, and its effect on the structure of the NS is investigated through the single fluid Tolman-Oppenheimer-Volkoff (TOV) formalism \cite{Lourenco:2022fmf, Lourenco:2021dvh, Lopes:2023uxi}. The quantity of DM trapped inside the star is controlled by adjusting the Fermi-momentum of the DM. The aim is to demonstrate that the XTE J1814-338 pulsar is most likely composed of a fermionic DM admixed NS.

{The work is organized as follows: In Sec.~\ref{fm} we present the formalism adopted for this work. Here we elaborate on the hadronic matter model in Sec.~\ref{hm}, the DM model in \ref{dm}, and the EoS parametrizations in \ref{pr}. In Sec.~\ref{mr}, we present and discuss the microscopic results, including the EoS, particle population, sound speed, and adiabatic index. In Sec.~\ref{ma} we present the macroscopic results, discussing the mass and radius relation and the moment of inertia, tidal deformability, and the Love number. In Sec.~\ref{fr} we present our comments and final remarks.
}

\section{Formalism}\label{fm}

\subsection{Hadronic matter}\label{hm}
To describe the interaction between the hadrons, we use quantum hadrodynamics (QHD)~\cite{Serot_1992}, where the strong force is simulated by the exchange of massive mesons. Considering the non-linear $\sigma\omega\rho\phi$ model, its Lagrangian density reads~\cite{IUFSU, Miyatsu2013}:

\begin{eqnarray}
\mathcal{L}_{QHD} = \bar{\psi}_B[\gamma^\mu(i\partial_\mu  - g_{B\omega}\omega_\mu   - g_{B\rho} \frac{1}{2}\vec{\tau} \cdot \vec{\rho}_\mu)  + \nonumber \\ 
- (M_B - g_{B\sigma}\sigma)]\psi_B  \nonumber  -U(\sigma) + \nonumber \\  
  + \frac{1}{2}(\partial_\mu \sigma \partial^\mu \sigma - m_s^2\sigma^2)+ \nonumber \\ - \frac{1}{4}\Omega^{\mu \nu}\Omega_{\mu \nu} + \frac{1}{2} m_v^2 \omega_\mu \omega^\mu + \frac{\xi g_{N\omega}^4}{4} (\omega_\mu\omega^\mu)^2 +\nonumber \\
 + \frac{1}{2} m_\rho^2 \vec{\rho}_\mu \cdot \vec{\rho}^{ \; \mu}  - \frac{1}{4}\bf{P}^{\mu \nu} \cdot \bf{P}_{\mu \nu} + \mathcal{L}_{\omega\rho} + \mathcal{L}_{\phi} , \label{s1} 
\end{eqnarray}
in natural units. $\psi_B$  is the baryonic  Dirac field, where $B$
can stand either for nucleons only ($N$) or can run over  nucleons ($N$) and hyperons ($Y$). The $\sigma$, $\omega_\mu$ and $\vec{\rho}_\mu$ are the mesonic fields, while $\vec{\tau}$ are the Pauli matrices.
 The $g's$ are the Yukawa coupling constants that simulate the strong interaction, $M_B$ is the baryon mass,  $m_s$, $m_v$, and $m_\rho$ are
 the masses of the $\sigma$, $\omega$ and $\rho$ mesons respectively.
The antisymmetric mesonic field tensors are given by their usual expression~\cite{Glenbook}: $\Omega_{\mu\nu} = \partial_{\mu}\omega_\nu - \partial_\nu\omega_\mu$, and $\bf{P}_{\mu\nu} = \partial_{\mu}\rho_\nu - \partial_\nu\rho_\mu$.
 The $\xi$ is related to the self-interaction of the $\omega$ meson while the $U(\sigma)$ is the scalar self-interaction term needed to correct the compressibility introduced in ref.~\cite{Boguta}:

  \begin{eqnarray}
U(\sigma) =  \frac{\kappa M_N (g_{N\sigma}\sigma)^3 }{3} + \frac{\lambda (g_{N\sigma}\sigma)^4}{4},\nonumber 
  \end{eqnarray}
 and $\mathcal{L}_{\omega\rho}$ is a non-linear
  $\omega$-$\rho$ coupling interaction as in ref.~\cite{IUFSU}:

\begin{eqnarray}
 \mathcal{L}_{\omega\rho} = \Lambda_{\omega\rho}(g_{N\rho}^2 \vec{\rho^\mu} \cdot \vec{\rho_\mu}) (g_{N\omega}^2 \omega^\mu \omega_\mu) , \label{EL2}
\end{eqnarray}

\noindent which is necessary to correct the slope of the symmetry
energy ($L$) and has a strong influence on the radii and tidal deformation of the neutron stars~\citep{Rafa2011,dex19jpg};
$\mathcal{L}_\phi$ is related  the strangeness hidden $\phi$ vector
meson, which couples only with the hyperons ($Y$), not affecting the
properties of symmetric  nuclear matter:

\begin{equation}
\mathcal{L}_\phi = g_{Y \phi}\bar{\psi}_Y(\gamma^\mu\phi_\mu)\psi_Y + \frac{1}{2}m_\phi^2\phi_\mu\phi^\mu - \frac{1}{4}\Phi^{\mu\nu}\Phi_{\mu\nu} , \label{EL3} 
\end{equation}
\noindent with $\Phi_{\mu\nu} = \partial_\mu \phi_\nu - \partial_\nu \phi_\mu$. As pointed in ref.~\cite{Lopes2020a,Weissnpa}, this vector channel is crucial to obtain massive hyperonic neutron stars. 
Applying the Euler-Lagrange formalism, and using the quantization rules ($E = \partial^0$, $k = i\partial^j$) we easily obtain the eigenvalue for the energy:

\begin{equation}
 E_B = \sqrt{k^2 + M_B^{*2}} + g_{B\omega}\omega_0 + g_{B\phi}\phi_0 + \frac{\tau_{3B}}{2}g_{B\rho}\rho_0 , \label{EL4}
\end{equation}
which at $T = 0$ K approximation also corresponds to the chemical potential.

As neutron stars are stable macroscopic objects, we need to describe a neutral, chemically
stable matter and hence, leptons are added as free Fermi gases. The EoS can now be obtained in mean-field approximation (MFA) by calculating the components of the energy-momentum tensor. The detailed calculation of the EoS in the mean-field approximation can be found
in refs.~\cite{Serot_1992,Miyatsu2013,Glenbook} and the references therein.

\subsection{Dark Matter}\label{dm}

The Lagrangian of the fermionic DM has a QHD-like form and reads~\cite{Lenzi:2022ypb,Lourenco:2021dvh,Das:2021hnk,Panotopoulos:2017idn,Lopes:2023uxi,Das:2020vng}:
\begin{eqnarray}
\mathcal{L}_{\rm DM} &=& \bar{\chi}(i \gamma^\mu \partial_\mu - (m_\chi -g_H h))\chi 
\nonumber \\
&&
+ \frac{1}{2}(\partial^\mu h \partial_\mu h - m_H^2 h^2). \label{FDMEOS}
\end{eqnarray}
Here, we assume a dark fermion represented by the Dirac field $\chi$ that self-interacts through the exchange of the Higgs boson, whose mass is $m_H$ = 125 GeV. The coupling constant is assumed to be $g_H = 0.1$,  which agrees with the constraints in Refs.~\cite{Panotopoulos:2017idn, Das:2021hnk}. In this framework, the DM self-interaction is extremely weak, resembling a tenuous, free Fermi gas. The dark matter energy eigenvalue is therefore:

\begin{equation}
 E_{\chi} = \sqrt{m_\chi^{*2} + k^2},   
\end{equation}
where $m_\chi^*$ = $m_\chi - g_Hh$ and $m_\chi$ = 200 GeV being the mass of the lightest neutralino as discussed in refs.~\cite{Lourenco:2022fmf, Lenzi:2022ypb, Das:2021hnk, Lopes:2023uxi}. Moreover, following those papers, we use the Fermi momentum to fix the dark matter content, up to $k_f = 0.08$ GeV.  As $k << m_\chi$, the pressure of the DM is almost zero.

As a matter of completeness, we also add a term that couples the baryonic matter with the DM as done in ref.~\cite{Das:2021hnk,Lenzi:2022ypb,Lourenco:2022fmf,Das:2020vng,Lourenco:2021dvh}.

\begin{equation}
\mathcal{L} = f\frac{M_B}{v} h\bar{\psi_B}\psi_B \label{cc}.
\end{equation}

Within this coupled channel, the effective mass of the baryon $B$ now also depends on the field $h$. In the same sense, the field $h$ now depends on both,  DM scalar density ($n_s^{DM}$) and the  baryonic scalar density $(n_s^B)$~\cite{Lenzi:2022ypb,Lourenco:2022fmf,Das:2020vng}:

\begin{eqnarray}
 M^*_B = M_B - g_{B\sigma}\sigma_0 -f\frac{M_B}{v}h ,  \nonumber \\
 h = \frac{g_H}{m_H^2}n_s^{DM} + \frac{f}{m_H^2}\sum_B\frac{M_B}{v}n_s^B ,
\end{eqnarray}
where $v$ = 246 GeV is the Higgs vacuum expectation value, and $f$ = 0.3 as done in ref.~\cite{Lenzi:2022ypb,Lourenco:2022fmf}. The total energy density and pressure are the sum of the hadronic, leptonic, and DM components { obtained through a mean-field approximation} (see ref.~\cite{Lenzi:2022ypb} and the references therein for a complete discussion).

\subsection{Parametrization}\label{pr}

It is worth noting that all results presented here are model-dependent. Therefore, the validity of the results strongly depends on whether the chosen parameterization fulfills the constraints coming from terrestrial nuclear laboratories and astrophysical observations. In the realm of nuclear physics, there are six parameters at the saturation density that must be fixed: the saturation density itself ($n_0$) the effective nucleon mass ($M_N^*/M_N$),  the binding energy per nucleon ($B/A$), the compressibility ($K$), the symmetry energy ($S_0$) and its slope ($L$).
The first four parameters are well known and were constrained in two review articles: refs.~\cite{Dutra2014,Micaela2017}. However, the symmetry energy and its slope are still a matter of debate, especially due to the inconsistencies between PREX and CREX experiments, as summarized in~\cite{Tagami2022}. Nevertheless, both were strongly constrained by combining astrophysical data with nuclear properties measured in the PREX-II experiment, together with chiral effective field theory, in Ref.~\cite{Essick2021}.

Regarding the astrophysical results, the NICER X-ray telescope has unambiguously established the existence of very massive neutron stars, with masses above two solar masses. 
We strongly believe that the main constraint that any realistic EOS must satisfy is the  PSR J0740+6620 pulsar whose mass and radius values
are $M = 2.08 \pm 0.07 M_\odot$ and $R =  12.39^{+1.30}_{-0.98}$ km~\cite{Miller2021}.
Another important constraint is the radius of the canonical
M = 1.4 M$_{\odot}$ star. We use here a very strong constraint
presented in Ref.~\cite{Riley2021}, that suggests $R_{1.4} = 12.45 \pm 0.65$ km, which limits the radius of the canonical star within an uncertainty of only 5$\%$. 
Ultimately, the LIGO/VIRGO gravitational wave observatories present constraints to the canonical star's dimensionless tidal parameter ($\Lambda$). An updated value for this quantity obtained from GW170817 is presented in ref.~\cite{AbbottPRL}, 70 $<~\Lambda_{1.4}<$ 580.

 \begin{table}[!t]
     \centering
     \begin{tabular}{c|c}
     \hline
     \multicolumn{2}{c}{L1$\omega^4$}
     \\ \hline
    $(g_{N\sigma}/m_s)^2$ & 11.800 fm$^2$ \\
    $(g_{N\omega}/m_v)^2$ & 6.945  fm$^2$ \\
    $(g_{N\rho}/m_\rho)^2$ & 9.00  fm$^2$ \\
    $\kappa$ & 0.00440 \\
    $\lambda$ &  $-0.0048$ \\
    $\Lambda_{\omega\rho}$ &  0.0530 \\ 
    $\xi$ & -0.00004 \\ \hline
     $(g_{H}/m_H)^2$ & 2.5$\times 10^{-8}$  fm$^2$  \\
     $(f/m_H)^2 \times (M_N/v)$ & 8.6 $\times 10^{-10}$ fm$^2$ \\ \hline
     \end{tabular}
     
\begin{tabular}{cc}
    { }&{ } \\
\end{tabular}

\begin{tabular}{c|cc}
\hline 
Quantity & Constraint & This model\\\hline
$n_0$ ($fm^{-3}$) & 0.148--0.170 & 0.164 \\
   $M^{*}/M$ & 0.60--0.80 & 0.69  \\

  $K$ (MeV)& 220--260  &  241  \\

 $S_0$ (MeV) & 31.2--35.0 &  33.1  \\
$L$ (MeV) & 38--67 & 44\\

 $B/A$ (MeV) & 15.8--16.5  & 16.0  \\ 
 \hline
\end{tabular}
 
\caption{ The L1$\omega^4$~\cite{lopes2024PRCb} parameterization (top) and its predictions to nuclear matter (bottom). The phenomenological constraints are taken from Ref.~\cite{Dutra2014, Micaela2017, Essick2021}. } 
\label{TL1}
\end{table}

A realistic EOS must therefore fulfill all the supracited constraints. To accomplish this task, we use a recent parameterization called L1$\omega^4$ presented in ref.~\cite{lopes2024PRCb}.
Based on previous studies about DM~\cite{Das:2020vng,Das:2021hnk,Lourenco:2021dvh,Lourenco:2022fmf}, changing the parametrization will not change the qualitative results. The parameters, the constraints, as well as the predicted physical quantities, are presented in Tab.~\ref{TL1}. The strength of the DM-Higgs boson interaction is also displayed in the same table, as well as the strength of the coupled channel $fM_N/v$. As can be seen, both are very feeble.
{Another interesting feature is the possible presence of hyperons in neutron stars' core. As the nucleons are subject to the Pauli exclusion principle, it can become energetically favorable at high densities to convert some of them into hyperons. However, the onset of hyperons softens the EOS, which can bring tension with the well-known existence of massive pulsars, such as the PSR J0740+6620. Such possible incoherence is called a hyperon puzzle.}

The possible presence of hyperons implies that we must fix their coupling constants with the scalar and vector mesons. To fix the hyperons-vector mesons coupling constants we assume that the Yukawa-Dirac Langrangian Eq.~({\ref{s1}}) is invariant under the flavor SU(3) symmetry group as done in ref.~\cite{lopesnpa,lopesPRD,lopes2023ptep}, while the interaction of the hyperons with the scalar meson is fixed in such a way that it reproduces the hyperons potential depth~\cite{Potentials2000,LQCD,Weissnpa}: $U_{\Lambda}$ = -28 MeV, $U_{\Sigma}$ = +30 MeV and $U_\Xi$ = -4 MeV. Moreover, to produce massive neutron stars even with a hyperon on their cores, we use here $\alpha_V$ = 0.25 (see ref.~\cite{lopesPRD} and references therein for a complete discussion about the relation of the coupling constants and the symmetry group). We then have:

 \begin{eqnarray}
   \frac{g_{\Lambda\omega}}{g_{N\omega}} &=&   0.75, \quad   \frac{g_{\Lambda\phi}}{g_{N\omega}}  = -1.06, \quad  \frac{g_{\Lambda\rho}}{g_{N\rho}} =0.0, \nonumber \\
   \frac{g_{\Sigma\omega}}{g_{N\omega}} &=& 1.25, \quad  \frac{g_{\Sigma\phi}}{g_{N\omega}} = -0.354, \quad \frac{g_{\Sigma\rho}}{g_{N\rho}}  = 0.5, \nonumber \\
   \frac{g_{\Xi\omega}}{g_{N\omega}} &=& 0.75, \quad  \frac{g_{\Xi\phi}}{g_{N\omega}} = -1.06, \quad \frac{g_{\Sigma\rho}}{g_{N\rho}}  = -0.5, \nonumber \\
   \frac{g_{\Sigma\Lambda\rho}}{g_{N\rho}} &=& 0.866, \label{couplings}
 \end{eqnarray}
 where we explicitly take into account the mixed-term 
$g_{\Sigma\Lambda\rho}$ 
to restore the relations of closure and completeness of the Clebsch-Gordon coefficients of the SU(3) group~\cite{lopes2023ptep}. 

 The study of DM-admixed neutron stars provides new constraints on the hyperon onset in dense matter. As it will become clear in the next sections, the presence of DM increases the central density of the maximally massive neutron star. Therefore, if we assume that hyperons are present in the absence of DM, they must also be present in DM-admixed NSs. Explicitly in the present study, if we assume that hyperons are present in the massive  PSR J0740+6620 pulsar, they must also be present in the XTE J1814-338 object.

\begin{figure*}[ht]
\begin{tabular}{ccc}
\centering 
\includegraphics[scale=.58, angle=270]{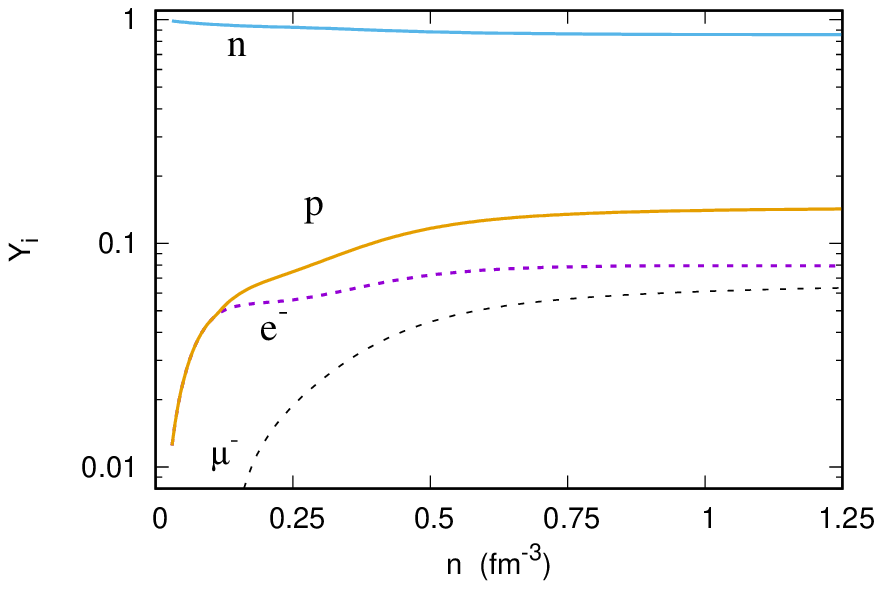} &
\includegraphics[scale=.58, angle=270]{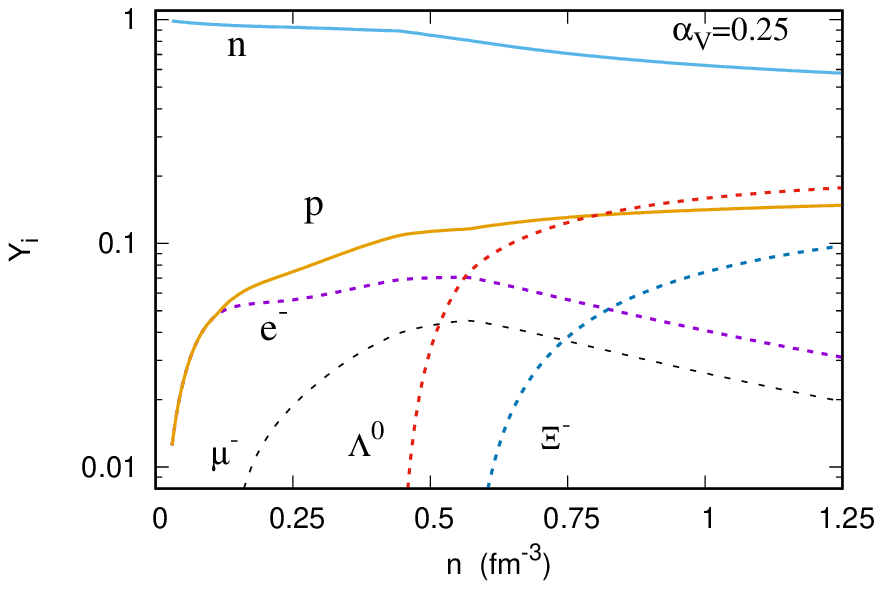} \\
\includegraphics[scale=.58, angle=270]{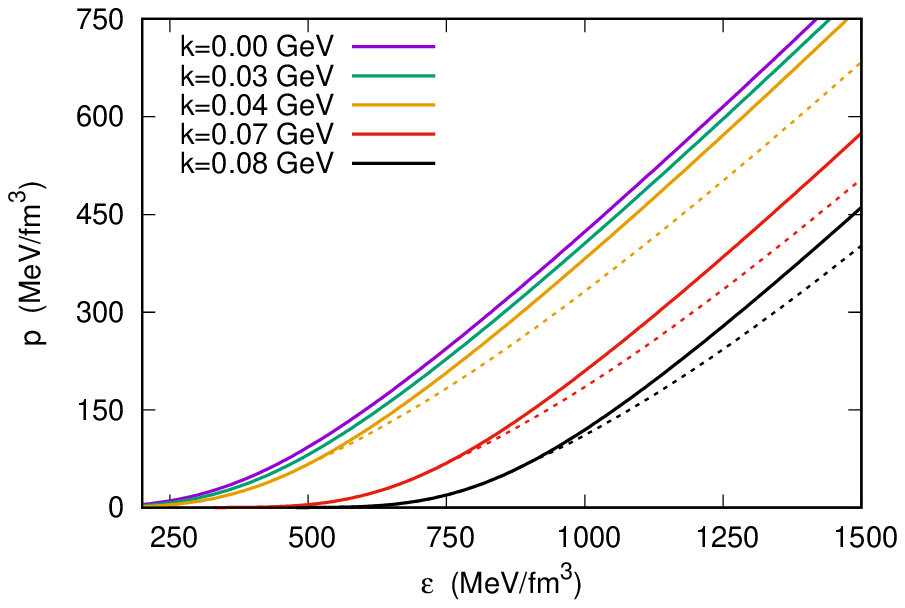} &
\includegraphics[scale=.58, angle=270]{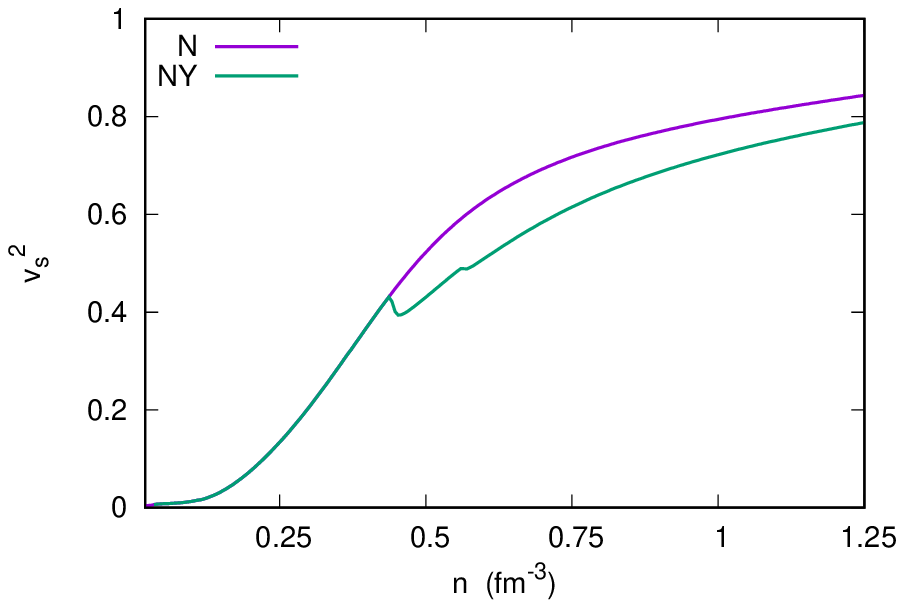}\\
\includegraphics[scale=.58, angle=270]{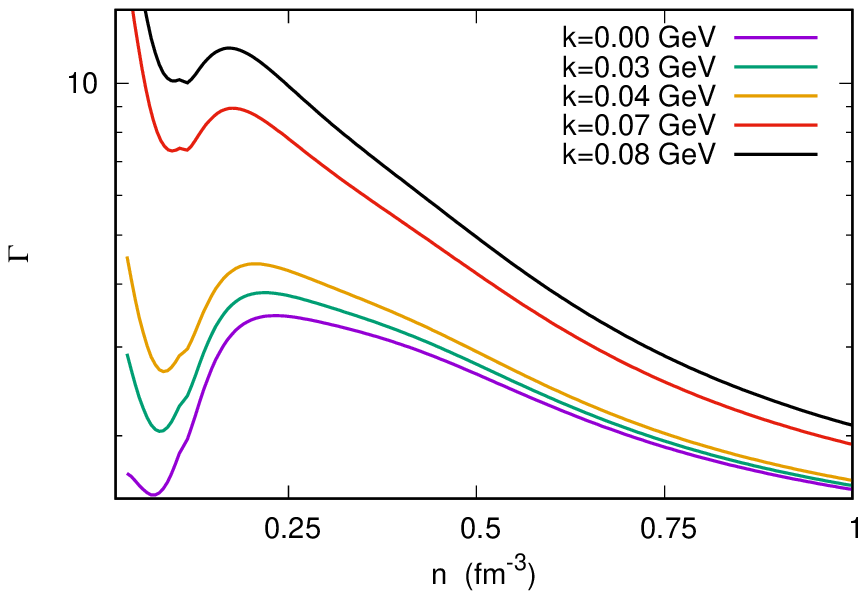} &
\includegraphics[scale=.58, angle=270]{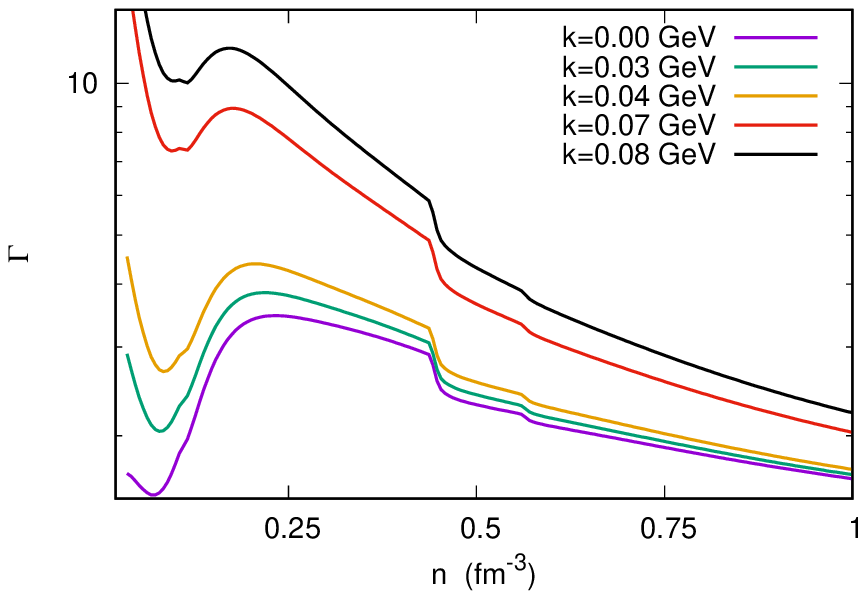}\\
\end{tabular}
\caption{Neutron stars' microscopic properties: Top: Particle population for nucleonic (left) and hyperonic (right) matter. Middle: EOS (left - dotted lines indicate hyperonic matter) and the speed of sound (right). Bottom: Adiabatic index $\Gamma$ for nucleonic (left) and hyperonic (right) matter.} \label{F1}
\end{figure*}

\section{Microscopic Results}\label{mr}

In this work, baryons and leptons are in chemical equilibrium, while the DM is introduced with different fixed Fermi momenta:  0.03 GeV, 0.04 GeV, 0.07 GeV, and 0.08 GeV; $k = 0.00$ GeV implies that there is no DM at all.  All the microscopic main properties are summarized in Fig~\ref{F1}.

At the top of Fig.~\ref{F1} we plot the particle population for nucleonic (left) and hyperonic  (right) matter. The DM is not shown as it enters with a fixed density. In the case of nucleonic matter, we can see that the fraction of electrons and muons grows with density. Moreover, the proton fraction is small and is below 0.15 even at high densities. In the case where we allow hyperons to appear in the stellar matter, we see that the first one that appears is the $\Lambda^0$ at 0.447 fm$^{-3}$, followed by the $\Xi^{-}$ at 0.570 fm$^{-3}$. 
We also notice a decrease in the leptons due to the onset of the negative charged $\Xi^-$ hyperon. The $\Xi^0$ and the $\Sigma$-triplet are absent within the chosen parametrization.

In the middle of Fig.~\ref{F1} we display the EOS (left) and the speed of sound (right) for different values of the DM Fermi moment, $k$. The EOSs that contain hyperons are dotted, while pure nucleonic EOS are shown as solid lines. While hyperons soften the EOSs at high densities, a constant DM inclusion softens the EOS as a whole. The softening is extreme for large values of $k$ as 0.07 and 0.08 GeV. On the other hand, the speed of sound $v_s^2 = \partial p /\partial \epsilon$ is not sensitive to the presence of DM. The reason is that this quantity is a derivative, while the contribution of DM enters as a constant in the energy density and the pressure. Nevertheless, hyperons can be seen in the speed of sound, as it causes a reduction in $v_s^2$ when both the $\Lambda^0$ and $\Xi^-$ appear.

A microscopic quantity that is sensitive to both hyperons and DM is the adiabatic index:

\begin{equation}
 \Gamma = \frac{(p+\epsilon)}{p} \bigg (\frac{\partial p}{\partial \epsilon} \bigg ). \label{add}
\end{equation}
For multicomponent matter the adiabatic index exhibits jumps at densities coincident with density
thresholds of individual components, signaling phase transitions, and/or changes in the matter constitution~\cite{lopesPRD}. The adiabatic index, $\Gamma$, presents information not only on
the EOS (p and $\epsilon$) but also on the speed of sound. The behavior of $\Gamma$ for nucleonic (left) and hyperonic (right) matter are displayed at the bottom of Fig.~\ref{F1} for different values of $k$.

Due to the extreme mass of the neutralino and the constant value of the energy density and pressure of the DM, $\Gamma$ is very large at low densities for large values of $k$, reaching $\Gamma > 10$ for densities below 0.1 fm$^{-3}$. Without DM, or for low values of $k$, $\Gamma$ has a maximum of around 0.22 fm$^{-3}$, however, this is not true for higher values of $k$. The presence of hyperons is easily recognized by the drop in $\Gamma$. The first one represents the onset of the $\Lambda^0$ and the more subtle drop represents the onset of the $\Xi^-$. 

\vspace{0.5cm}

\section{Macroscopic Results}\label{ma}

\begin{figure*}[ht]
\begin{tabular}{ccc}
\centering 
\includegraphics[scale=.58, angle=270]{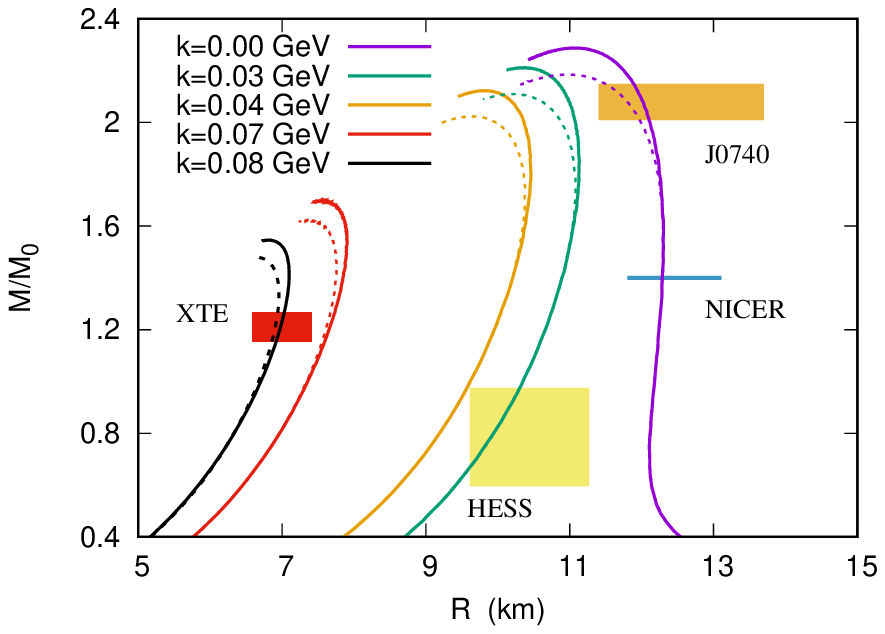} &
\includegraphics[scale=.58, angle=270]{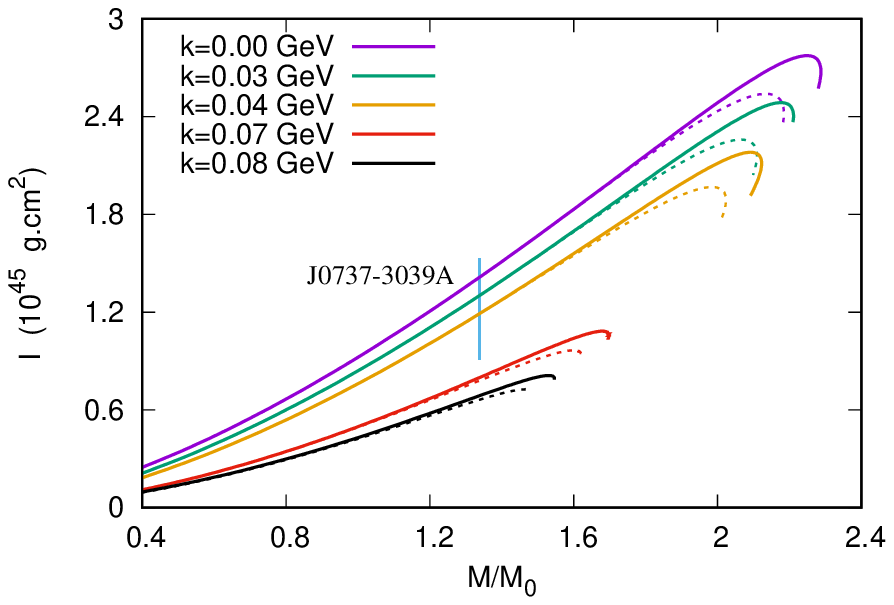} \\
\includegraphics[scale=.58, angle=270]{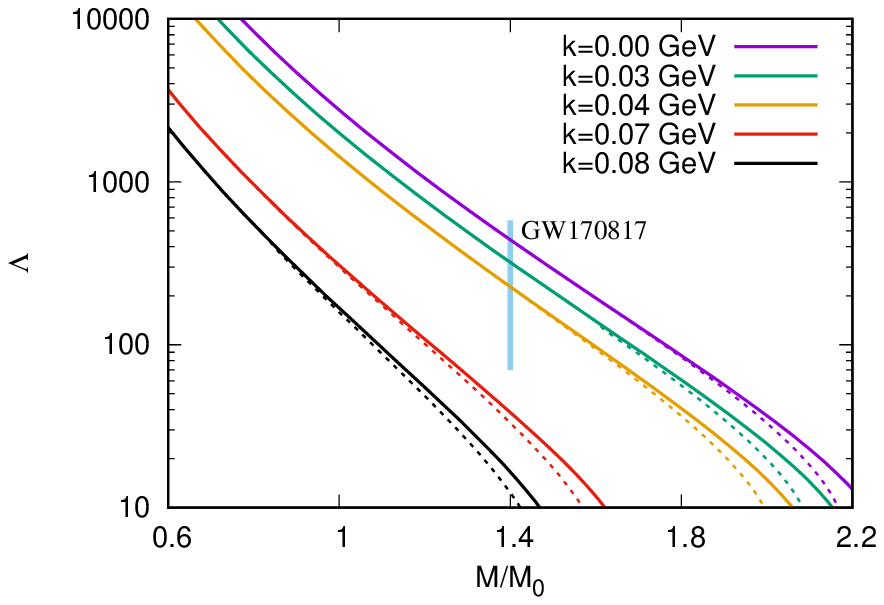} &
\includegraphics[scale=.58, angle=270]{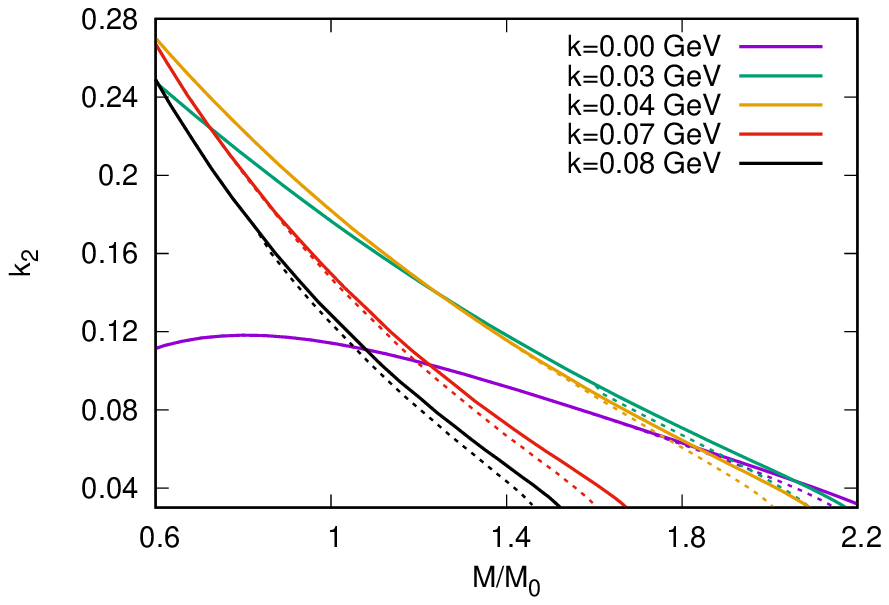}\\
\end{tabular}
\caption{Neutron star's macroscopic properties. Top: Mass-radius relation (left) and the MOI (right). Bottom: The dimensionless tidal parameter $\Lambda$ (left) and the Love number $k_2$ (right). Dotted lines imply hyperons in the core while solid lines represent pure nucleonic neutron stars.} \label{F2}
\end{figure*}

We now discuss the macroscopic features of DM-admixed neutron stars. Our first analysis is the mass-radius relation, obtained by solving the TOV equations~\cite{TOV}:

\begin{eqnarray}
 \frac{dp}{dr} = \frac{-GM(r)\epsilon (r)}{r^{2}} \bigg [ 1 + \frac{p(r)}{\epsilon(r)} \bigg ]   \bigg  [ 1 + \frac{4\pi p(r)r^3}{M(r)} \bigg ] \nonumber \\ \times \bigg [ 1 - \frac{2GM(r)}{r} \bigg ]^{-1} , \nonumber \\
 \frac{dM}{dr} =  4\pi r^2 \epsilon(r).  \nonumber \\  \label{Etov}
\end{eqnarray}

The EOS is therefore used as an input to solve the Eqs.~(\ref{Etov}). To describe the outer and inner crusts of the neutron stars, we utilized the Baym-Pethick-Sutherland (BPS) EOS \cite{BPS} and the Baym-Bethe-Pethick (BBP) EOS \cite{BBP}, respectively.   We use the BPS+BBP EoS up to the density of 0.0089 fm$^{-3}$ for all values of $k$, and from this point on, we use the QHD EOS, as suggested in ref.~\cite{Glenbook}. This procedure is the same as the one done in ref.~\cite{lopesPRD,lopes2024PRCb,lopesnpa,lopes2023ptep,Lopes2022ApJ}, but here we also take into account the DM contribution. As the dark matter enters as a constant in the energy density (as well as in the pressure but with several orders of magnitude below), $\epsilon$ does not go to zero at the stars' surfaces. Since we are dealing with one fluid approach, all stars are stable up to the maximum mass~\cite{Glenbook}.

In the top left of Fig.~\ref{F2}, we plot the mass-radius relation along with some observational constraints for different values of $k$. Pure nucleonic neutron stars are described as solid lines, while dotted lines represent hyperonic neutron stars.  As we already pointed out, maybe the more important constraint is the mass and radius measurement of the  PSR J0740+6620 pulsar, $M = 2.08 \pm 0.07 M_\odot$ and $R =  12.39^{+1.30}_{-0.98}$ km~\cite{Miller2021}. For the canonical M = 1.4 M$_{\odot}$ star we use here the constraint presented in Ref.~\cite{Riley2021}, $R_{1.4} = 12.45 \pm 0.65$ km. A standard model (without DM) must be able to satisfy both constraints simultaneously.

In this work, two very compact objects are analyzed. The first is the HESS J1731-347~\cite{Doroshenko_2022}, whose mass and radius are $M=0.77_{-0.17}^{+0.20}~M_\odot$ and $R = 10.4 _{-0.78}^{+0.86}$ km respectively. Finally, the key point of the present work, the XTE J1814-338, with an inferred mass and radius of $M = 1.21 \pm 0.05 M_\odot$ and R = 7.0 $\pm$ 0.4 km~\cite{Kini:2024ggu}, presents a big challenge to modern nuclear astrophysics.

As can be seen from left-top Fig.~\ref{F2}, without DM, the canonical and the  PSR J0740+6620 pulsar can be simulated with the L1$\omega^4$ even in the presence of hyperons. Indeed, the radius of the canonical star within the chosen model is 12.30 km. Nevertheless, the HESS J1731-347 and the XTE J1814-338 can only be described (within the present parametrization) as a DM-admixed neutron star. In the case of the HESS object, a small amount of DM ($k = 0.03$ GeV) is enough to reproduce the inferred mass and radius.
However, XTE J1814-338 can only be reproduced with a significantly larger amount of DM ($k = 0.08 $ GeV). Such a large value for the Fermi momentum of the neutralino exceeds what is typically considered in the literature. For instance, in Refs.~\cite{Lenzi:2022ypb, Lourenco:2021dvh, Lourenco:2022fmf, Lopes:2023uxi, Das:2018frc}, the neutralino Fermi momentum was not considered to exceed 0.06 GeV. Moreover, simultaneously describing the PSR J0740+6620 and the XTE J1814-338  with a single EOS seems very unlikely. This fact strongly suggests they must have a different nature.

We now discuss how the presence of DM affects the maximum mass and the central baryonic density. Without DM we have a maximum mass of 2.29 $M_\odot$ with a central density of 0.95 fm$^{-3}$ for a pure nucleonic star and a maximum mass of 2.18 $M_\odot$ with a central density of 1.01 fm$^{-3}$ for a hyperonic neutron star. For  $k$ = 0.03 GeV, we have 2.21 $M_\odot$ with 0.97 $fm^{-3}$ for nucleonic and 2.11 $M_\odot$ with 1.09 fm$^{-3}$ for hyperonic star. For the extreme case of $k$ = 0.08 GeV, we obtain a maximum mass of 1.55 $M_\odot$ ($1.48~M_\odot$) with a central density of 1.48 fm$^{-3}$ (1.62 fm$^{-3}$) for a nucleonic (hypersonic) neutron star. We can see that the presence of DM reduces the masses and the radii of the stars while causing a significant increase in the central baryonic density.

We also analyze the $M_{min}$, that is, the minimum neutron star mass that presents hyperons in their core as a function of $k$. For a non-DM neutron star, hyperons are only present for stars with $M~>1.5M_\odot$. For $k =0.003$ GeV, we have $M_{min} = 1.37M_\odot$. For the extreme case, $k$ = 0.08 GeV, we have $M_{\min} = 0.45M_\odot$. This implies that XTE J1814-338 can also present hyperons in its core.

We now analyze the moment of inertia (MOI) of the stars for different values of $k$ with and without hyperons. For a static neutron star, the moment of inertia can be calculated as~\cite{Glenbook}:

\begin{equation}
  I = \frac{8}{3}\int_0^R r^4 \frac{(\epsilon(r) +p(r))}{\sqrt{1-2GM(r)/r}} \cdot e^{-\nu} dr, \label{inertiam}  
\end{equation}
with

\begin{equation}
  \frac{d \nu}{dr} = G\frac{M(r) +4\pi r^3 p(r)}{r(r -2GM(r))}  .
\end{equation}

It is important to point out that there is no observational measure of the moment of inertia of neutron stars. Nevertheless, the MOI of the PSR J0737-3039A was estimated as $1.15^{+0.38}_{-0.24}~\times~10^{45}$ g cm$^2$ based on universal relations analyses~\cite{Landry_2018}, with a well-measured mass of 1.338 $M_\odot$.
For non-DM admixed neutron stars, the L1$\omega^4$ predicts an MOI of 1.40 $\times 10^{45}$  g.cm$^2$, for a 1.338 $M_\odot$ neutron star, therefore in agreement with the bound from the PSR J0737-3039A. The results and this constraint are presented in the top-right of Fig.~\ref{F2}.

We can therefore estimate the MOI of some specific objects. For instance, without DM, the canonical 1.4$M_\odot$ has an MOI of 1.51 $\times 10^{45}$ g.cm$^2$. A 2.08 $M_\odot$ pulsar, corresponding to the PSR J0740+6620 has an MOI of 2.60 $\times 10^{45}$ g.cm$^2$ for a nucleonic neutron star and 2.51 $\times 10^{45}$ g.cm$^2$ for a hyperonic star. The HESS J1731-347 object can be obtained with $k = 0.03$ GeV. In this case, for a mass of 0.77 $M_\odot$, we have an MOI of 0.57 $\times 10^{45}$ g.cm$^2$. Finally, the XTE J1814-338, which can only be obtained using $k$ =0.08 GeV has an MOI of  0.58 $\times 10^{45}$ g.cm$^2$ considering it a nucleonic neutron star and  0.56 $\times 10^{45}$ g.cm$^2$ if we consider it as a hyperonic one. Although significantly different in mass and radius, the HESS J1731-347 and the XTE J1814-338, have a very similar MOI.

Ultimately, we discuss the dimensionless tidal parameter $\Lambda$.  If we put an extended body in an inhomogeneous external field, it will experience different forces throughout its surface. The result is a tidal interaction. The tidal deformability of a compact object is a single parameter $\lambda$ that quantifies how easily the object is deformed when subjected to an external tidal field. Larger tidal deformability indicates that the object is easily deformed. Conversely, a compact object with a small tidal deformability parameter is more compact and more difficult to deform. The tidal deformability is defined as:
\begin{equation}
 \Lambda~\equiv~\frac{\lambda}{M^5}~\equiv~\frac{2k_2}{3C^5} , \label{tidal}
\end{equation}
where $M$ is the mass of the compact object and $C = GM/R$ is its compactness. The parameter $k_2$ is called the second (order) Love number. Additional discussion about the theory of tidal deformability and the tidal Love numbers is beyond this work's scope and can be found in Refs.~\cite{Abbott2017, AbbottPRL, Abbott:2018wiz, Flores2020, tidal2010, Lourenco:2021dvh, Lopes2022ApJ,Lopes:2023uxi}.  Nevertheless, as pointed out in Refs.~\cite{tidal2010, Lourenco:2021dvh}, the value of $y_R$ must be corrected when a discontinuity is present.
This is the case of DM-admixed neutron stars, once they have a finite energy density on the surface, as they are taken into account with a fixed Fermi moment. Therefore, we must have
\begin{equation}
 y_R \rightarrow y_R - \frac{4\pi R^3 \Delta\epsilon_S}{M} , \label{yr}
\end{equation}
where $R$ and $M$ are the star radius and mass, respectively, and $\Delta\epsilon_S$ is the difference between the energy density at the surface ($p=0$) and the star's exterior (which implies $\epsilon=0$). The results for the dimensionless tidal parameter altogether with the constraint, 70 $<~\Lambda_{1.4}~<$ 580~\cite{AbbottPRL}; and the Love number $k_2$ are displayed at the bottom of Fig.~\ref{F2}.

We first notice that, with $\Lambda_{1.4}$ = 446, the L1$\omega^4$ fulfill the constraint for the GW170817. As already pointed out in ref.~\cite{lopes2024PRCb}, there is a competition between the compactness $C$ and the Love number $k_2$. In general, for a fixed parametrization, reducing the slope of the symmetry energy causes a reduction of the neutron stars' radii and consequently a reduction in the compactness $C$, however, at the same time, it causes an increase in the $k_2$.  Analogously here, the addition of DM causes a reduction in $C$ but an increase in $k_2$. This is even more evident here because of the discontinuity of $y_R$, which causes $k_2$ to reach high values for low-mass neutron stars. Nevertheless, the dimensionless tidal parameter always decreases with $k$. A complete summary of the results presented in this work is presented in tab.~\ref{T2}.

\begin{widetext}
\begin{center}
\begin{table}[ht]
\begin{center}
\scriptsize
\begin{tabular}{cc|cccccccccccccc}
\hline
  $k$ (GeV) & Model &$M_{\mathrm{max}} (M_\odot)$ &  $n_c$ (fm$^{-3}$) & $R_{1.4}$ (km) & $I_{0.77}$ (g.cm$^2$)  &$I_{1.2}$ (g.cm$^2$) & $I_{1.4}$ (g.cm$^2$) & $I_{2.08}$ (g.cm$^2$)& $\Lambda_{0.77}$  & $\Lambda_{1.2}$ &$\Lambda_{1.4}$ & $\Lambda_{2.08}$ & $M_{min}~(M_\odot)$  \\
\toprule
 0.00 & N & 2.29 &  0.95 &12.30 & 0.63 $\times~10^{45}$ & 1.21 $\times~10^{45}$  & 1.51 $\times~10^{45}$  & 2.60 $\times~10^{45}$  & 10495 &  1071 & 446 & 27 & - \\
 0.00 &NY & 2.18 & 1.01 &12.30 & 0.63 $\times~10^{45}$  & 1.21 $\times~10^{45}$  & 1.51 $\times~10^{45}$  & 2.51 $\times~10^{45}$  & 10495&  1071 & 446 & 22 & 1.50 \\
0.03 &N & 2.21 & 0.97 & 10.90 & 0.57 $\times~10^{45}$  & 1.11 $\times~10^{45}$  & 1.40 $\times~10^{45}$  & 2.41 $\times~10^{45}$  & 7277 &  783& 327 &17 & - \\ 
0.03 &NY & 2.11 & 1.09 &10.90 & 0.57 $\times~10^{45}$ & 1.11 $\times~10^{45}$ & 1.40 $\times~10^{45}$  & 2.25 $\times~10^{45}$  & 7277 &  783 &327 & 11 & 1.37 \\
0.04 &N & 2.12 & 1.05 & 10.20 & 0.51 $\times~10^{45}$   & 1.01 $\times~10^{45}$  & 1.28 $\times~10^{45}$   & 2.18 $\times~10^{45}$  & 5148 &  559 & 231 & 8 & - \\
  0.04& NY &2.02 &  1.11 & 10.19 & 0.51 $\times~10^{45}$  & 1.01 $\times~10^{45}$  & 1.28 $\times~10^{45}$  & -   &  5148 & 559 &  229 & - & 1.22 \\
 0.07&N &1.70 & 1.41 & 7.85& 0.32 $\times~10^{45}$ &0.67 $\times~10^{45}$  & 0.86 $\times~10^{45}$  &  - & 1173   &  112 & 39 &  - & -  \\
 0.07 &NY&1.62 & 1.45 & 7.75 & 0.32 $\times~10^{45}$ & 0.66 $\times~10^{45}$  & 0.83 $\times~10^{45}$  & -  & 1173   &  105 & 34 &  - & 0.63  \\
 0.08 &N &1.55 & 1.48 & 7.10 & 0.27 $\times~10^{45}$  & 0.58 $\times~10^{45}$ & 0.74 $\times~10^{45}$ & - & 678   &  57 & 17 & - & -  \\
0.08 & NY &1.48 & 1.62  & 6.93 & 0.27 $\times~10^{45}$  & 0.56 $\times~10^{45}$  & 0.70 $\times~10^{45}$  & -  &  678 &  50 & 13 & - &  0.45  \\
\hline
\end{tabular}
\caption{Neutron stars' main properties for different values of $k$. } \label{T2}
\end{center}
\end{table}
\end{center}
\end{widetext}

\section{Final Remarks}\label{fr}

In this work, we study the influence of DM in neutron stars' interiors, focusing on the ultra-compact object XTE J1814-338 pulsar. We show that such an object can be explained as a DM admixed neutron star with a high value for the DM Fermi moment, $k$ = 0.08 GeV. We also show that this object can even present hyperons in its core. Additional results are summarized below.

\begin{itemize}

    \item We use the L1$\omega^4$~\cite{lopes2024PRCb} parametrization of the QHD. This model agrees with all six constraints that come from nuclear physics at saturation density~\cite{Dutra2014,Micaela2017,Essick2021}, as well as from nuclear astrophysics, as inferred by the NICER relative to the canonical 1.4~$M_\odot$~\cite{Riley2021} and the 2.08 $M_\odot$ PSR J0740+6620 pulsar~\cite{Miller2021}. Furthermore, it also satisfies the bound related to the MOI of the PSR J0737-3039A~\cite{Landry_2018} and the dimensionless tidal parameter inferred by the LIGO/VIRGO in the GW170817 event~\cite{AbbottPRL}.

    \item The presence of DM as well as the presence of hyperons softens the EOS. However, the softening due to DM has a bigger effect.  On the other hand, the speed of sound is insensible to the presence of DM once it is added as a constant in the total energy density and pressure. Nevertheless, the adiabatic index is sensible to both, the DM and hyperons. Indeed, for high values of $k$ we can have $\Gamma~>$ 10 at low densities because the presence of dark matter introduces a jump in the value of $\Gamma$. Thus, the greater the value of $k$, the higher the dark matter energy density, and the more pronounced the observed jump.

    \item The presence of hyperons and DM compacts the star, but only DM can compact neutron stars with low masses. Hyperons and DM also increase the central density and reduce the maximum mass. 

    \item The HESS J1731-347 object can be modeled as a DM admixed neutron star with a small Fermi momentum (0.03 GeV).  However,  the XTE J1814-338 needs a much higher value (0.08 GeV). Despite presenting different values of radius and mass, both objects share a similar MOI-.
    It is very unlikely that the XTE J1814-338 and the PSR J0740-6620 objects have the same nature.

    \item The presence of DM increases the Love number $k_2$ for low-mass neutron stars, mainly due to the discontinuity in $y_R$. Nevertheless, the dimensionless tidal parameter always decreases with $k$. Consequently, the presence of DM alters the internal and physical structure of the star, thereby affecting its ability to resist external tidal forces.

    \item Finally, the discrepancies in the mass, radius, and tidal deformability of neutron stars, compared to the known observational predictions, potentially caused by the presence of dark matter, could serve as indirect evidence for the detection of DM in neutron stars.

\end{itemize}

\textbf{Acknowledgements:} L.L.L. was partially supported by CNPq Universal Grant No. 409029/2021-1. A.I. would like to acknowledge the financial support from the São Paulo State Research Foundation (FAPESP) under Grant No. 2023/09545-1. Both authors thank Prof. Debora Peres Menezes for the fruitful discussion that helped to improve the manuscript.

\bibliography{aref}

\end{document}